\documentclass[aps,prl,twocolumn,groupedaddress]{revtex4}
\usepackage{amssymb,amsmath}
\usepackage{epsfig}
\usepackage{graphicx}

\begin{document}

\title{Staggered-Vortex Superfluid of Ultracold Bosons in an Optical Lattice}
\author{Lih-King Lim and C. Morais Smith}
\affiliation{Institute for Theoretical Physics, Utrecht University, 3508 TD Utrecht, The Netherlands}
\author{Andreas Hemmerich}
\affiliation{Institute f\"{u}r Laser-Physik,
Universit\"{a}t Hamburg, Luruper Chaussee 149, 22761 Hamburg, Germany}
\date{\today}

\begin{abstract}
We show that the dynamics of cold bosonic atoms in a two-dimensional square optical lattice produced by a bichromatic light-shift potential is described by a Bose-Hubbard model with an additional effective staggered magnetic field. In addition to the known uniform superfluid and Mott insulating phases, the zero-temperature phase diagram exhibits a novel kind of finite-momentum superfluid phase, characterized by a quantized staggered rotational flux. An extension for fermionic atoms leads to an anisotropic Dirac spectrum, which is relevant to graphene and high-$T_c$ superconductors.
\end{abstract}

\maketitle
\date{\today}
The experimental realization of ultracold atomic gases loaded into optical lattices has opened up a unique pathway to study quantum phase transitions of many-body systems \cite{BH_Model}. Exploiting the rich internal atomic structure and the great versatility in the engineering of optical potentials, increasingly complex optical lattice models have been proposed \cite{Buc:05, OL&Unconven}, for which unconventional quantum phases are predicted. Recently, it has been suggested that optical lattices can be used to simulate the quantum behavior of charged particles in a two-dimensional (2D) lattice subjected to a homogeneous magnetic field \cite{OL&QuantumHall}. This system is known to exhibit a wealth of interesting physics, such as the famous Hofstadter butterfly single-particle spectrum \cite{Hof:76} or the integer and fractional quantum Hall effects \cite{QuantumHall}. Much less is known, however, about charged particles moving in a 2D lattice subjected to a {\it staggered} magnetic field. In a pioneering work, Haldane has shown that the integer quantum Hall effect may occur in this system as a result of broken time-reversal symmetry \cite{Hal:88}. Concerning the single-particle spectrum, only recently numerical studies have revealed the connection to the Hofstadter butterfly \cite{Wan:06}. The technical difficulty to engineer magnetic fields alternating on the spatial scale of condensed matter lattices has constrained experimental studies to magnetic fields modulated on a mesoscopic scale \cite{Mesoscopic}.
\begin{figure}
\includegraphics[scale=.29, angle=0, origin=c]{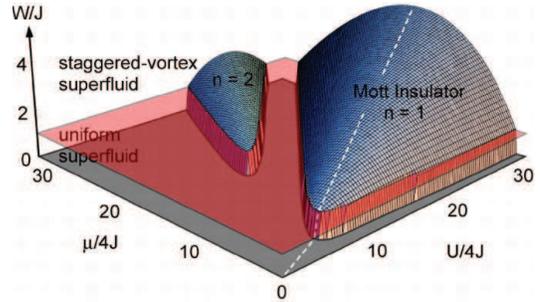}
\caption{\label{Fig.1} (color online). Phase diagram with respect to the chemical potential $\mu$, the interaction parameter $U$, the hopping amplitude $J$, and the scaled magnetic flux $W$. Within the three-dimensional lobes a gapped Mott insulator phase prevails. Outside these lobes the system is superfluid. For small magnetic fields ($W/J < 1$) the superfluid is spatially uniform, while for large magnetic fields ($W/J > 1$), a quantized staggered rotational flux arises. The white dashed line indicates the ($\mu/U = 2-\sqrt{2}$) plane analyzed in detail in Fig.~\ref{Fig.2}.}
\end{figure}

\begin{figure}
\includegraphics[scale=.35, angle=0, origin=c]{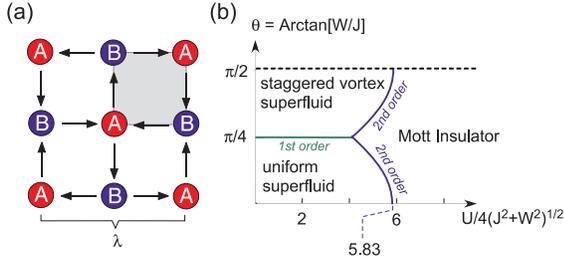}
\caption{\label{Fig.2}(color online). (a) The bipartite lattice comprises sites labelled by ``A'' and ``B.'' The gray rectangle identifies a single plaquette. The solid arrows indicate the tunnelling currents driven by $V_R(\mathbf{r},t)$. (b) Phase diagram within the ($\mu/U = 2-\sqrt{2}$)-plane spanned by the dashed white line and the W axis in Fig.~\ref{Fig.1}.}
\end{figure}

In this letter we show that a staggered magnetic field in a lattice can be realized for ultracold bosonic and fermionic atoms in a 2D optical lattice. We then present a mean field theory for the bosonic system and construct the zero-temperature phase diagram shown in Fig.~\ref{Fig.1}. For small magnetic fields, the spatially uniform $(k = 0)$ superfluid phase and the Mott insulating state, known from the conventional Bose-Hubbard model, are reproduced. When the magnetic flux per plaquette $\Phi$ exceeds $\Phi_0 /2$ (see Fig.~\ref{Fig.2}(a) for the definition of plaquette), where $\Phi_0=hc/e$ is the fundamental flux quantum, we find that a novel finite momentum $(k = \pi)$ superfluid phase is realized.
This superfluid phase is spatially modulated and is characterized by quantized fluxes of alternating sign for adjacent plaquettes.
Our work thus points out the possibility to realize a novel spatially modulated superfluid with ultracold bosons, which exhibits a characteristic momentum spectrum. An extension of our work for fermions offers the possibility to simulate various strongly correlated systems, such as the mean-field Hamiltonian of Affleck and Marston \cite{MAffleck88}, proposed in the context of high-$T_c$ superconductors\cite{Wen:06}, and the case of massless Dirac spectra, as realized in graphene \cite{Neto07}.

\textit{Effective Hamiltonian.}--We first consider bosonic atoms trapped in a 2D square optical lattice potential $V_L(\mathbf{r}) \equiv - V_0(\sin^2(kx)+\sin^2(ky))$ with an additional time-dependent potential $V_R(\mathbf{r},t) \equiv \kappa V_L(\mathbf{r})\cos(2S(\mathbf{r})-\Omega t)$, $S(\mathbf{r})\equiv\tan^{-1}[ (\sin(kx)-\sin(ky))/( \sin(kx)+\sin(ky))]$, which acts as a collection of micro-rotors, each applying angular momentum to a single plaquette, with alternating signs for adjacent plaquettes. Here, $k\equiv2\pi/\lambda$ and $\lambda$ is the wavelength of the optical potential. The term $V_R(\mathbf{r},t)$ can be implemented experimentally by means of a bichromatic light-shift potential \cite{Hem:07}. The well depth $V_0 > 0$, the oscillation frequency $\Omega$ and the coupling strength $\kappa\in[0,1]$ are adjustable parameters in experiments. We denote the sublattices of the bipartite square lattice by A and B (cf. Fig.~\ref{Fig.2}(a)) and define four vectors $\mathbf{e}_1=-\mathbf{e}_3=(\lambda/2)\hat{x}$, $\mathbf{e}_2=-\mathbf{e}_4=(\lambda/2)\hat{y}$ connecting each A site to its four B neighbors. Upon the assumption that the atoms in the optical lattice are restricted to the lowest Bloch band and that $V_R(\mathbf{r},t)$ does not induce interband transitions, the system can be described by a Bose-Hubbard model with time-varying hopping and energy offset \cite{BH_Model, Hem:07}
\begin{eqnarray}
\label{Hamiltonian}
\hat H(t)= -\negthickspace\sum_{\mathbf{r}\in A, l=1-4} \negthickspace J_l(t)\, \left\{\hat{a}^{\dag}_{\mathbf{r}} \,\, \hat{a}_{\mathbf{r}+\mathbf{e}_l}+ \textrm{H.c.}\right\}
\nonumber\\
+\sum_{\mathbf{r}\in A\oplus B} \epsilon_{\mathbf{r}}(t) \,\, \hat{n}_{\mathbf{r}}+\frac{1}{2}\,U\negthickspace\negthickspace\sum_{\mathbf{r}\in A\oplus B}\hat{n}_{\mathbf{r}}\,\,(\hat{n}_{\mathbf{r}}-1)\, ,
\end{eqnarray}
where $\hat{a}_{\mathbf{r}}$ and $\hat{a}^{\dag}_{\mathbf{r}}$ are the boson annihilation and creation operators on site $\mathbf{r}$ obeying the canonical commutation relation $[a_{\mathbf{r}},a^{\dag}_{\mathbf{r}'}] =\delta_{\mathbf{r},\mathbf{r}'}$, $\hat{n}_{\mathbf{r}}=\hat{a}^{\dag}_{\mathbf{r}}\, \hat{a}_{\mathbf{r}}$ is the number operator, $J_l(t)=J+(-1)^l\kappa V_0 \chi_1 \sin(\Omega t)$ denotes the anisotropic time-varying hopping, $\epsilon_{\mathbf{r}\in A,B}(t) = \pm2\kappa V_0 \chi_2 \cos(\Omega t)$ is a time-varying energy offset, and $U$ is the onsite interaction energy. In terms of the Wannier function of the lowest band $w(\mathbf{r})$, we have $J = \int dxdy \, w^*(x+\lambda/4,y) [-\frac{\hbar^2}{2m}\nabla^2+V_L(\mathbf{r})]w(x-\lambda/4,y)$, $\chi_1 = \int dxdy \, w^*(x+\lambda/4,y)[\sin^2(kx)-\cos^2(ky)]w(x-\lambda/4,y)$, $\chi_2 = \int dxdy \, |w(x,y)|^2 [\cos(kx)\cos(ky)]$ and $U \propto (a_s/m) \int dxdy \, |w(x,y)|^4$, with the atomic mass $m$ and the $s$-wave scattering length $a_s$. We note that by using Feshbach resonances, the $s$-wave interaction can be considered as another experimentally tunable parameter.
Upon expressing all terms with harmonic time-dependence in terms of a quantized auxiliary bosonic field and integrating out this field, similarly as in Ref.~\cite{Buc:05}, we find the effective Hamiltonian
\begin{eqnarray}
\label{Ham1}
\hat{H}_{\textrm{eff}}\approx -\negthickspace\sum_{\mathbf{r} \in A,l=1-4} \left\{|c|{\rm e}^{i\theta(-1)^l}\hat{a}_{\mathbf{r}}^{\dag} \,\, \hat{a}_{\mathbf{r}+\mathbf{e}_l}
+\textrm{H.c.}\right\}\nonumber\\+\frac{1}{2}U\negthickspace\negthickspace\sum_{\mathbf{r}\in A\oplus B} \negthickspace\hat{n}_{\mathbf{r}}\,(\hat{n}_{\mathbf{r}}-1),
\end{eqnarray}
where the anisotropic complex hopping amplitude is given by $|c|=\sqrt{J^2+W^2}$, $\theta=\tan^{-1}(W/ J)$ with $W=2\kappa^2V_0^2\chi_1 \chi_2/\hbar \Omega$. The Aharonov-Bohm phase $\theta$, picked up by the bosons when tunnelling along the edge of a plaquette, may be interpreted as resulting from an effective staggered magnetic field with a magnetic flux per plaquette $\Phi=(2\theta / \pi )\Phi_0$. This artificial gauge field arises due to the $\pi/2$ phase lag between the time-varying hopping terms $J_l(t)$ and the energy offset terms $\epsilon_{\mathbf{r}\in A,B}(t)$, which assigns a sense of rotation to each plaquette. In Eq.~(\ref{Ham1}) we have neglected non-local negative ring exchange terms with energy $(\kappa V_0 \chi_1)^2/2 \hbar\Omega$ which may be tuned to be 2 orders of magnitude smaller than all other terms. We also omitted a nonlocal term composed of summands $\hat{n}_\mathbf{r}\hat{n}_\mathbf{r'}$, describing interaction between distant lattice sites mediated by the retroaction of the atoms upon the light-shift potential. This term is expected to be significant, if the light-shift potential is produced inside a high-finesse optical cavity \cite{Mas:05}.  For conventional light-shift potentials produced by superposition of laser beams, which we consider here, it is irrelevant. Nonetheless, this term does not alter the mean-field results presented below.

\begin{figure}
\includegraphics[scale=.3, angle=0, origin=c]{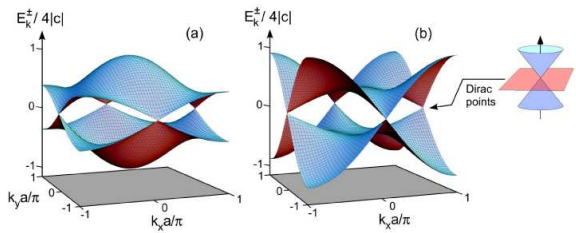}
\caption{\label{Fig.3} (color online). Single-particle spectrum for (a) $\theta<\pi/4$, with the minimum at the origin, and (b) $\theta>\pi/4$, with minima at the Brillouin zone edges $(\pm\pi/a,\pm\pi/a)$.}
\end{figure}

For an optical lattice loaded with fermionic atoms, a similar derivation leads to an effective staggered magnetic field as well. The single-particle spectrum can be obtained for both, fermions and bosons, by expressing the kinetic terms in Hamiltonian (\ref{Ham1}) in momentum space with two independent amplitudes
$\hat{a}_{\mathbf{k}}=\Sigma_{\mathbf{r}\in A} \,\hat{a}_{\mathbf{r}} \,e^{i
\mathbf{k}\cdot \mathbf{r}}, \hat{b}_{\mathbf{k}}=\Sigma_{\mathbf{r}\in B}\,
\hat{a}_{\mathbf{r}}\, e^{i\mathbf{k}\cdot \mathbf{r}}$, and applying the canonical
transformations
$\hat{\alpha}_{\mathbf{k}}=[(\epsilon^*_{\mathbf{k}}/|\epsilon_{\mathbf{k}}|)
\hat{a}_{\mathbf{k}} +\hat{b}_{\mathbf{k}}]/\sqrt{2},
\,\,\hat{\beta}_{\mathbf{k}}=[-(\epsilon^*_{\mathbf{k}}/|\epsilon_{\mathbf{k}}|)
\hat{a}_{\mathbf{k}} +\hat{b}_{\mathbf{k}}]/\sqrt{2}$. Here, $\epsilon_{\mathbf{k}}=2|c|e^{i\theta}\cos[k^+ a]+2|c|e^{-i\theta}\cos[k^- a]$ is the generalized lattice dispersion, with $k^{\pm}=(k_x\pm k_y)/2$ and $a=\lambda/\sqrt{2}$. We then find
\begin{eqnarray}
E^{\pm}_{\mathbf{k}}&=&\pm2|c|[\cos^2\left(k^+ a\right)+
\cos^2\left(k^- a\right)\nonumber\\&&+2\cos\left(k^+ a\right)
\cos\left(k^- a\right)\cos\left(2\theta\right)]^{1/2},
\end{eqnarray}
where the lower (upper) branch corresponds to the in-(out-of-)phase mode
$\hat{\beta}_{\mathbf{k}}$ ($\hat{\alpha}_{\mathbf{k}} $) between the sublattices. For the fermionic system at half-filling, the single-particle spectrum exhibits two inequivalent anisotropic Dirac cones; see Fig.~\ref{Fig.3}. Their slope can be tuned via the staggered magnetic field, thus resembling the physics of graphene with anisotropic hopping arising under uniaxial pressure \cite{Hasegawa06}. Furthermore, at $\theta=\pi/4$ and negligible interactions, the system simulates the Affleck-Marston Hamiltonian \cite{MAffleck88}, wherein a staggered $\pi$-flux phase is proposed to describe the pseudogap regime of high-$T_c$ superconductors. Although several experiments were suggested to probe the relevance of this phase, the problem remained open due to unavoidable effects of disorder \cite{cha01}. The system we consider thus offers an opportunity to study such phases in a highly controllable environment.

\textit{Mean-field theory for the bosonic system.}--We anticipate the bosonic system to exhibit the well-known quantum phase transition from a superfluid to a Mott insulating phase as the parameter $U/4|c|$ is increased \cite{BH_Model}. The phase boundary between the two phases is determined by a straightforward generalization of the functional integration method presented in Ref. \cite{Oos:01}. The partition function is written in terms of an imaginary-time functional integral $Z=\int \!\mathcal{D}a^*\mathcal{D}a \exp \{-S[a^*,a]/\hbar\}$ with the action $S[a^*,a]=\int_0^{\hbar/k_B T} d \tau\left[ \sum_{\mathbf{r}} a_{\mathbf{r}}^*(\tau) \left(\hbar\partial_\tau -\mu\right)a_{\mathbf{r}}(\tau)+ H_{\textrm{eff}} \right]$, where $T$ is the temperature and $\mu$ the chemical potential. In the Mott regime, we introduce the order parameter $\psi_{\mathbf{r}}(\tau)$ (a Hubbard-Stratonovich field) to decouple the hopping term and upon integrating out the boson fields $(a^*,a)$, the effective action (up to quadratic order) gives the zero-temperature quasiparticle(hole) energy dispersion,
\begin{eqnarray}
\epsilon^{qp,qh}_{\mathbf{k}}=-\mu+\frac{U}{2}(2n-1)-\frac{|\epsilon_{\mathbf{k}}|}{2}\pm\frac{1}{2}\hbar
\, \omega_{\mathbf{k}},
\end{eqnarray}
where $\hbar
\omega_{\mathbf{k}}=\sqrt{|\epsilon_{\mathbf{k}}|^2-(4n+2)|\epsilon_{\mathbf{k}}|U+U^2}$ is the energy required for creating a quasiparticle-quasihole pair and $n$ is the lattice filling factor, given by the ratio between the number of bosons $N$ and the number of sites $N_s$. The phase boundary between the Mott and the superfluid states can then be determined by the condition that the excitation becomes gapless $\hbar \omega_{\mathbf{k}}=0$, which is shown in Fig.~\ref{Fig.1} as the generalized Mott lobes. For brevity we henceforth concentrate on the ($\mu / U = 2-\sqrt{2}$)-plane spanned by the dashed white line and the W-axis in Fig.~\ref{Fig.1} [cf. Fig.~\ref{Fig.2}(b)]. We note that for $\theta>\pi/4$ the low energy excitations in the gapped Mott phase carry a finite, rather than zero, lattice momentum. This implies that the associated critical phenomena at the phase boundary between the Mott-insulator and the staggered vortex phase could be different from the usual universality class of the $O(2)$ quantum rotor model for the superfluid-Mott transition in the Bose-Hubbard model \cite{Sac:99}.

In the superfluid regime, however, the staggered magnetic field drives the system into distinct superfluid phases. Even though interactions induce a finite quantum depletion, for a system of $N$ weakly interacting bosons, Bose-Einstein condensation (BEC) takes place at the lowest single-particle state. For $\theta<\pi/4$, BEC occurs at the $\mathbf{k}=(0,0)\equiv 0$ state, giving rise to a uniform superfluid with the many-body ground state $|\Psi_{0}\rangle\propto(\hat{\beta}^\dag_0)^N|0\rangle=(\sum_{\mathbf{r}\in A\oplus B}\hat{a}^{\dag}_{\mathbf{r}})^N|0\rangle$. For $\theta>\pi/4$, new absolute minima develop at the Brillouin zone edges $\mathbf{k}=(\pm\pi/a,\pm\pi/a)\equiv\pi$ for which condensation takes place; see Fig.~\ref{Fig.3}. Because of the equivalence of the four minima in the reciprocal space, the new many-body ground state can be written as
$|\Psi_{\pi}\rangle\propto(\hat{\beta}^\dag_\pi)^N|0\rangle = (\sum_{\mathbf{r}} \, (\hat{a}_{\mathbf{r}}^{\dag} + i \hat{a}_{\mathbf{r}+\mathbf{e}_2}^{\dag} - \hat{a}_{\mathbf{r}+\mathbf{e}_1+\mathbf{e}_2}^{\dag} - i \hat{a}_{\mathbf{r}+\mathbf{e}_1}^{\dag}))^N |0\rangle$ with $ \mathbf{r} = 2\mathbf{r'}$ and $\mathbf{r'}\in A\oplus B $. The angular phases of the order parameter differ by $\pi/2$ between neighboring lattice points, and there is a quantized flux on each plaquette, with alternating sign for adjacent plaquettes. Thus, for the BEC at $\mathbf{k}=\pi$, the system is characterized by a vortex-antivortex lattice with a periodicity $\lambda/\sqrt{2}$, namely, a staggered-vortex superfluid phase. This phase possesses a definite chirality that is commensurate with the staggered magnetic field, a feature which is not present in the uniform superfluid phase. To confirm that distinct superfluids are stabilized, we employ a variational mean-field ansatz for the ground state $|\xi,\sigma\rangle=(e^{-i\xi/2} \cos(\sigma) \hat{\beta}^\dag_0+ e^{i\xi/2} \sin(\sigma) \hat{\beta}^\dag_\pi )^N|0\rangle $ and minimize the expectation value with respect to the Hamiltonian (\ref{Ham1}). The uniform superfluid ($\sigma=\sigma_0=0$) and the staggered-vortex phase ($\sigma=\sigma_0=\pi/2$) are indeed the absolute minima of the mean-field energy for $\theta<\pi/4$ and $\theta>\pi/4$, respectively. Furthermore, the stability of both phases can be verified by examining the energy cost for deviating from the condensate $\langle \hat{H} \rangle_{|\sigma_0+\varepsilon\rangle}=E_{MF}+\varepsilon^2 4 N|c|\bigr[\sqrt{2}|\sin(\pi/4-\theta)|+(U N )/(8|c|)\bigl]+ {\cal O}(\varepsilon^4)$ with $\sigma_0 \in \{0,\,\pi/2\}$ and $E_{MF}=-4N|c|\cos(\theta)$. Finally, we note that as the system is tuned across the $\theta=\pi/4$ line, $\sigma_0$ changes discontinuously by a value of $\pi/2$, suggesting that the two superfluid phases are separated by a quantum first-order phase transition line within this variational mean-field analysis. At this line, both phases are degenerate in their mean-field energies, with a finite energy barrier $\Delta\sim U N^2/8 N_s$ between the two minima.

\textit{Fluctuations.}--Following the Bogoliubov theory for a weakly interacting Bose gas, we derive the energy spectra of the superfluid phases. We consider the Hamiltonian (\ref{Ham1}) in the grand-canonical ensemble and make the substitution for the condensation mode $\hat{\beta}_{\mathbf{k}_0}\rightarrow\sqrt{N_0}+\hat{\beta}_{\mathbf{k}_0}$, where $N_0$ is the condensate number and $\mathbf{k}_0=0$ ($\mathbf{k}_0=\pi$) for the uniform superfluid (staggered-vortex) phase. Choosing the chemical potential at its mean-field value, $\mu=-|\epsilon_{\mathbf{k}_0}|+n_0U/2$ and keeping terms up to quadratic order in the fluctuations, the Hamiltonian becomes,
\begin{eqnarray*} \nonumber
\hat{H}_{\mathbf{k}_0}&\approx&-\frac{1}{4}n_0U N_0+\sum_{\mathbf{k}}\left\{
\left(|\epsilon_{\mathbf{k}}|-|\epsilon_{\mathbf{k}_0}|+\frac{1}{2}n_0
U\right)\hat{\alpha}^{\dag}_{\mathbf{k}}
\hat{\alpha}_{\mathbf{k}}\right.\\&&
\left.+\left(-|\epsilon_{\mathbf{k}}|-|\epsilon_{\mathbf{k}_0}|+\frac{1}{2}n_0
U\right)\hat{\beta}^{\dag}_{\mathbf{k}}
\hat{\beta}_{\mathbf{k}}\right.\\&&
\left.+\left[\,\frac{1}{8}n_0U
A_{\mathbf{k},\mathbf{k}_0} (\hat{\alpha}_{\mathbf{k}}^\dag\hat{\alpha}_{\mathbf{-k}}^\dag+
\hat{\beta}_{\mathbf{k}}^\dag\hat{\beta}_{\mathbf{-k}}^\dag) \right.\right.
\\&&\left.\left.+\frac{1}{4}n_0UB_{\mathbf{k},\mathbf{k}_0}\hat{\alpha}_{\mathbf{k}}^\dag\hat{\beta}_{\mathbf{-k}}^\dag+\textrm{H.c.} \right]\right\},
\end{eqnarray*}
where $n_0=N_0/N_s$ is the condensate density, $A_{\mathbf{k},0}=B_{\mathbf{k},\pi}=1+\exp(-2i\varphi_\mathbf{k})$, $B_{\mathbf{k},0}=A_{\mathbf{k},\pi}=1-\exp(-2i\varphi_\mathbf{k})$, and $\varphi_\mathbf{k}=\arg (\epsilon_{\mathbf{k}})$. The Hamiltonian can be readily diagonalized by a Bogoliubov transformation to yield the spectrum $\hbar\omega_{\mathbf{k},\mathbf{k}_0}=\sqrt{|\epsilon_\mathbf{k}|^2+|\epsilon_{\mathbf{k}_0}|^2+n_0U|\epsilon_{\mathbf{k}_0}|\pm n_0U|\epsilon_\mathbf{k}| \sqrt{G_{\mathbf{k},\mathbf{k}_0}}}$, where $G_{\mathbf{k},0}=\cos^2 ( \varphi_\mathbf{k} )+4 |\epsilon_0|[|\epsilon_0|/n_0U+1]/n_0U$ and $G_{\mathbf{k},\pi}=\sin^2 (\varphi_\mathbf{k} )+4 |\epsilon_{\pi}|[|\epsilon_{\pi}|/n_0U+1]/n_0U$. The lower branch of the spectrum is linear and gapless at long wavelength $\hbar \omega_{\mathbf{k},\mathbf{k}_0}\approx \sqrt{|c|\cos(\theta-k_0 a/2)[4|c|\cos(\theta-k_0 a/2) +n_0 U]} \,\,(\mathbf{k}-\mathbf{k}_0)$ corresponding to the Goldstone mode of the broken gauge symmetry.

\begin{figure}
\includegraphics[scale=.19, angle=0, origin=c]{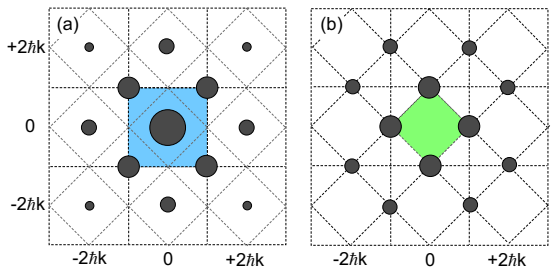}
\caption{\label{Fig.4}(color online). Schematic of the momentum spectra for the uniform superfluid (a) and the staggered-vortex superfluid (b). The colored areas in the centres illustrate the first Brillouin zones with (b) and without (a) the staggered magnetic field.}
\end{figure}

\textit{Experimental signatures.}--The characteristic momentum spectrum of the staggered-vortex phase provides a clear signature to identify this state experimentally by imaging momentum space using standard ballistic expansion techniques. The momentum distribution can be expressed as $\langle \Psi^{\dag}(\mathbf{k}) \Psi(\mathbf{k})\rangle\,=\,|W(\mathbf{k})|^2 \,\,  S_B(\mathbf{k})\, S_P(\mathbf{k})$, where $W(\mathbf{k})$ is the Fourier transform of the Wannier function, $S_B(\mathbf{k}) = |\sum_{\mathbf{R}\in \textrm{A}\oplus \textrm{B}} \textrm{e}^{i 2 \mathbf{k}\cdot\mathbf{R}}|^2$ is the structure factor of the Bravais lattice, and $S_P(\mathbf{k}) = \sum_{\nu,\mu \in \{1,2,3,4\}} \textrm{e}^{i \mathbf{k}\cdot(\mathbf{r}_{\nu}-\mathbf{r}_{\mu})} \langle \hat{a}_{\mathbf{r}_{\nu}}^{\dag}\hat{a}_{\mathbf{r}_{\mu}}\rangle$ is the structure factor of a plaquette. Here, $\mathbf{r}_{\nu}$ denote the four corners of a plaquette and $\hat{a}_{\mathbf{r}_{\nu}}$ are the corresponding boson operators. We may write $S_P(\mathbf{k}) = n \, |\sum_{\nu \in \{1,2,3,4\}} \textrm{e}^{i \mathbf{k}\cdot\mathbf{r}_{\nu}} \textrm{e}^{i \psi_{\nu}}|^2$ with $\psi_{\nu} = 0$ for the uniform superfluid $|\Psi_{0}\rangle$ and $\psi_{\nu} = \nu \pi /2$ for the staggered vortex superfluid $|\Psi_{\pi}\rangle$. As illustrated in Fig.~\ref{Fig.4}, the two cases display distinct structures of Bragg maxima, directly observable in experiments.

In conclusion, we have shown that anisotropic and time-varying hopping terms in a 2D optical lattice give rise to an effective staggered magnetic field. For the bosonic system, it leads to a novel kind of superfluid phase characterized by a quantized staggered rotational flux. For the system realized with fermionic atoms, it gives rise to anisotropic Dirac spectra at half-filling. The tunability of the interaction terms and the addition of optical disorder potentials allow for systematic simulations of various strongly correlated systems, such as graphene and high-$T_c$ superconductors. Another exciting direction for future work could be the search for quantum Hall physics in the system. Finally, we remark that the experimentally accessible Hamiltonian (\ref{Hamiltonian}) offers a wider parameter space than presently considered in this work. The inclusion of non-negligible ring exchange interactions, for example, may offer the opportunity to realize exotic quantum insulators \cite{Buc:05}.

\begin{acknowledgments}
A. H. acknowledges support by DFG (He2334/10-1). We would like to thank R. Duine, \mbox{M. P. A. Fisher}, R. Moessner, and \mbox{H. T. C. Stoof} for fruitful discussions.
\end{acknowledgments}

\end{document}